\documentclass[12pt]{article}
\usepackage{amssymb,amsmath,epsfig}
\allowdisplaybreaks

\begin{document}

\title{\bf Lorentz Distributed Noncommutative Wormhole Solutions in Extended Teleparallel Gravity}
\author{Abdul Jawad \thanks {abduljawad@ciitlahore.edu.pk, jawadab181@yahoo.com} and
Shamaila Rani \thanks {drshamailarani@ciitlahore.edu.pk, shamailatoor.math@yahoo.com}\\
Department of Mathematics, COMSATS Institute of\\ Information
Technology, Lahore, Pakistan.}

\date{}

\maketitle
\begin{abstract}
In this paper, we study static spherically symmetric wormhole
solutions in extended teleparallel gravity with the inclusion of
noncommutative geometry under Lorentzian distribution. We obtain
expressions of matter components for non-diagonal tetrad. The
effective energy-momentum tensor leads to the violation of energy
conditions which impose condition on the normal matter to satisfy
these conditions. We explore the noncommutative wormhole solutions
by assuming a viable power-law $f(T)$ and shape function models. For
the first model, we discuss two cases in which one leads to
teleparallel gravity and other is for $f(T)$ gravity. The normal
matter violates the weak energy condition for first case while there
exist a possibility for micro physically acceptable wormhole
solution. There exist a physically acceptable wormhole solution for
the power-law $b(r)$ model. Also, we check the equilibrium condition
for these solutions which is only satisfied for teleparallel case
while for $f(T)$ case, these solutions are less stable.
\end{abstract}

\section{Introduction}

General theory of relativity laid down the foundation of modern
cosmology. The $\Lambda$CDM model is the simplest model compatible
with all cosmological observations but suffers some shortcomings
like fine-tuning and cosmic coincidence problems. The modified
theories of gravity are the generalized models came into being by
modifying gravitational part in general relativity (GR) action while
matter part remains unchanged. At large distances, these modified
theories, modify the dynamics of the universe. In another scenario,
modified matter part with unchanged gravitational part results
dynamical models such as cosmological constant, quintessence,
k-essence, Chaplygin gas and HDE models \cite{S26,FS1}. After GR,
Einstein attempted to unify gravitation and electromagnetism based
on mathematical structure of absolute or distant parallelism, also
referred as teleparallelism which led to teleparallel (TP) gravity.
In this gravity, the gravitational field is established through
torsion using Witzenb\"{o}ck connection instead of curvature via
Levi-Civita connection in GR.

The extended TP theory of gravity (or $f(T)$ gravity where $T$
represents torsion scalar) is the generalization of TP gravity in
the same fashion as $f(R)$ gravity generalizes GR. This is attained
by replacing torsion scalar in the action of TP gravity by a general
differentiable function $f(T)$. Ferraro and Fiorini \cite{S1}
firstly introduced this theory by applying Born-Infeld strategy and
solved particle horizon problem as well as obtained singularity-free
solutions with positive cosmological constant. Afterwards, this
theory has been studied extensively under many phenomena, like
accelerated expansion of the universe, solar system constraints,
discussion of Birkhoff's theorem, thermodynamics, reconstruction via
dynamical models, static and dynamical wormhole solutions, viability
of models through cosmographic technique, instability ranges of
collapsing stars, and many more. A vast area of research in $f(T)$
gravity is fulfilled using spherically symmetric scenario
\cite{8th}.

A wormhole is a hypothetical path to connect different regions of
the universe which can be regarded as a tunnel or bridge from which
an observer may traverse easily. Wormhole solutions are described by
static as well as dynamical configurations. In GR, the exotic matter
(violates the energy condition) constitutes basic ingredient to
develop mathematical structure of wormhole. The violation of NEC is
the necessary tool to form wormhole solutions which also allow two
way travel. Also, the inclusion of electromagnetic field, scalar
field, noncommutative (NC) geometry, NC Lorentzian (NCL)
distribution, \cite{S301} etc establish more interesting and useful
results. The search for a realistic source which provides the
violation (while normal matter satisfies the energy conditions) has
recently gained a lot of interest. The modified theories of gravity
are one of the direction to establish realistic wormhole solutions.

In $f(T)$ gravity, the effective energy-momentum tensor is the cause
for the corresponding violation while normal matter threads wormhole
solutions. In this regard, the extra terms related to the underlying
theory play their effective role to violate the energy conditions
which is necessary to keep wormhole throat open to be traversable.
B\"{o}hmer et al. \cite{S15} were the first who studied wormhole
solutions in this gravity by taking static spherically symmetric
traversable wormhole solutions and found some constraints on the
wormhole throat. They explored behavior of energy conditions by
taking particular $f(T)$, shape and redshift functions and obtained
physically acceptable solutions. Jamil et al. \cite{S16} studied
these solutions by taking fluid as isotropic, anisotropic as well as
particular equation of state and found that energy conditions are
violated in anisotropic case while these are satisfied for the
remaining two cases. Sharif and Rani \cite{10'} explored dynamical
wormhole solutions for traceless as well as barotropic equations of
state. With the help of analytic and numerical $f(T)$ models, they
concluded that weak energy condition (WEC) holds in specific time
intervals for these cases.

Rahaman et al. \cite{20c3} examined the NC wormhole solutions in GR
for higher dimensional static spherically symmetric spacetime and
found their existence in regular way upto four dimensions while in a
very restrictive way for five dimensional space. Beyond these
dimensions, the possibility of wormhole solutions is over. Jamil et
al. \cite{22c3} also explored $f(R)$ wormhole solutions in the same
background. Recently, Sharif and Rani have studied wormhole
solutions \cite{11} in the background of NC geometry for
$f(T)=\alpha T^n$ model as well as shape function. They concluded
that the effective energy-momentum tensor serves as the basic
ingredient to thread the wormhole solutions and normal matter gives
some physically acceptable solutions. They extended \cite{11a} this
work to study the effects of electrostatic field. The same authors
explored these solutions for galactic halo regions \cite{asd} for
exponential and logarithmic models but no physically acceptable
solutions are obtained.

Recently, Bhar and Rahaman \cite{fr} studied the wormhole solutions
in Lorentzian distribution as the density function in the
noncommutativity-inspired spacetime. They obtained a stable wormhole
which is asymptotically flat in the usual four dimensional
spacetime. In order to search for some realistic wormhole solutions,
we extended this work in extended teleparallel gravity for specific
$f(T)$ and shape function models. The paper is organize as follows:
In the next section, we provide wormhole geometry for static
spherically symmetric spacetime and briefly discuss the energy
conditions. Section \textbf{3} is devoted to the discussion of
$f(T)$ gravity. In section \textbf{4}, we construct the field
equations and matter content for the wormhole solutions. Section
\textbf{5} contains the construction of wormhole solutions for
particular $f(T)$ and $b(r)$ functions. In section \textbf{6}, we
check the equilibrium condition for the obtained solutions. The last
section summarizes the results.

\section{Wormhole Geometry}

One of the most fascinating features of GR is the possible existence
of spacetimes with non-trivial topological structure. The well-known
examples of this structure are described by Misner and Wheeler
\cite{S34} and Wheeler \cite{S35} as solutions of the Einstein field
equations known as wormhole solutions. Basically, the wormhole
having appearance as tube, tunnel or bridge represents a shortcut
way to communicate between two distant regions. If there exist some
other paths between these regions then these are called
``intra-universe", otherwise ``inter-universe" wormholes. The
simplest example of such connection is the Einstein-Rosen bridge (or
Schwarzschild wormhole) which unfortunately fails to provide a way
of communication to another region of space to which it is
connected. The reason behind this is that any photon or particle
falling in it, reaches the singularity at $r=0$ and thus has no link
with other end of wormhole.

The Lorentzian traversable wormholes are the most favorable in the
sense that a human may traverse from one side of the wormhole to the
other through these wormholes. Being the generalized form of
Schwarzschild wormhole (having event horizon which permits one way
travel) with no event horizon, these wormholes lead to two way
travel. Morris and Thorne \cite{S36} established static spherically
symmetric wormhole spacetime given by
\begin{equation}\label{126a}
ds^2=e^{2\Psi(r)}dt^2-\frac{dr^2}{1-\frac{b(r)}{r}}-r^2d\theta^2-r^2\sin^2\theta
d\phi^2.
\end{equation}
Here, $\Psi(r)$ represents redshift (or potential) function which
determines gravitational redshift of a light particle (photon). The
magnitude of redshift function must be finite everywhere for the two
way travel through wormhole. The function $b(r)$ denotes the shape
function which sets the shape of the wormhole.

The essential characteristics required for a wormhole geometry are
discussed in \cite{S36,S37} given as follows.
\begin{itemize}
\item The no-horizon condition ($\Psi(r)$ must be
finite everywhere in the spacetime) must be satisfied by the
redshift function. Usually, it is taken as zero, which gives
$e^{2\Psi(r)}\rightarrow 1$.
\item The shape function must satisfy the flaring out condition on
throat, i.e., to have the proper shape for a wormhole, the ratio of
the radial coordinate to the shape function at that coordinate must
be $1$ while this coordinate expresses non-monotonic behavior away
from throat. Taking throat radius as $r_0$, it yields $b(r_0)=r_0$
and $b'(r_0)<1$.
\item The proper radial distance,
$\tau(r)=\pm \int^{r}_{r_0} \frac{dr}{\sqrt{1-b(r)/r}},~r\geq r_0$
should be finite throughout the space. The signs $\pm$ associate the
two parts which are joined by the throat for wormhole configuration.
\item At large distances, the asymptotic flatness should be accomplished by the
spacetime, i.e., $\frac{b(r)}{r} \rightarrow 0$ as
$r\rightarrow\infty$.
\end{itemize}
Notice that Morris-Thorne wormhole is not a particular wormhole
solution like Schwarzschild wormhole which depends on a single
parameter, the mass of wormhole. Instead, it is a class of solutions
for arbitrarily large number of redshift as well as shape functions
satisfying the above constraints.

In order to prevent shrinking of wormhole throat and to make it
traversable, the matter distribution of energy-momentum tensor at
throat must be negative. More precisely, the sum of energy density
and pressure of matter is negative representing violation of Null
energy condition (NEC) and such matter is named as ``exotic". It is
noted that ordinary energy-momentum tensor satisfies the NEC. For
viability of wormhole solutions, it is necessary to minimize the
amount of exotic matter required to support wormhole solutions. The
modified theories of gravity are one of the source which provide
effective energy-momentum tensor to violate the WEC (contains NEC).
In this regard, the usual energy-momentum tensor may satisfy this
condition. To discuss NEC and WEC, we assume energy-momentum tensor
in appropriate orthonormal frame as
\begin{equation*}
\mathcal{T}^{\alpha\beta}=\textmd{diag}(\rho,~p_1,~p_2,~p_3),
\end{equation*}
where $\rho$ is the energy density and $p_n (n=1,2,3)$ denote
principal pressures.
\begin{itemize}
\item \textbf{Weak Energy Condition:} The relationship between Raychaudhuri equation
and attractiveness of gravity yields the WEC as
\begin{equation*}
\mathcal{T}_{\alpha\beta}V^\alpha V^\beta\geq0,
\end{equation*}
for any timelike vector $V^\alpha$. In terms of components of the
energy-momentum tensor, this inequality yields
\begin{equation*}
\rho\geq0,\quad \rho+p_n\geq0,\quad \forall\quad n.
\end{equation*}
For modified theories of gravity, we replace matter content of the
universe in effective manner as $\mathcal{T}^{eff}_{\alpha\beta}$ as
well as matter components $\rho^{_{eff}}$ and $p_n^{_{eff}}$.
\item \textbf{Null Energy Condition:} By continuity, the WEC implies the NEC
\begin{equation*}
\mathcal{T}_{\alpha\beta}k^\alpha k^\beta\geq0,
\end{equation*}
for any null vector $k^\alpha$. This inequality gives
$\rho+p_n\geq0,~\forall~n$. In effective manner, this becomes
$\rho^{_{eff}}+p_n^{_{eff}}\geq0$.
\end{itemize}

\section{$f(T)$ Theory of Gravity}

It is well-known that curvature and torsion are the properties of a
connection and many connections may be defined on the same
spacetime. The Riemann-Cartan (generalized Riemannian) spacetime
yields two interesting models of spacetime such as Riemannian and
Weitzenb\"{o}ck spacetimes \cite{S30}. The concept of curved
manifold is a crucial characteristic of GR which is carried out
through Riemannian spacetime having metric tensor as the dynamical
variable. It uses curvature (Riemann) tensor to achieve Levi-Civita
connection whose curvature remains non-zero while torsion vanishes
due to its symmetry property. On the other hand, the non-zero
torsion with vanishing curvature corresponds to the Weitzenb\"{o}ck
spacetime which parallel transports the tetrad field. The TP theory
is defined by Weitzenb\"{o}ck connection with tetrad field as the
basic entity. This theory is an alternative description of gravity
having translation group which is related to a gauge theory. The
existence of non-trivial tetrad field in gauge theory leads to the
teleparallelism structure. The $f(T)$ theory of gravity is the
generalization of TP theory.

The geometry of TP theory is uniquely defined by an orthonormal set
of four-vector fields (three spacelike and one timelike) named as
the tetrad field. The simplest type of tetrad field is the trivial
tetrad which has the form
$e_i=\delta_{i}^\mu\partial_\mu,~e^j=\delta^j_\mu dx^\mu$, where
$\delta^i_\mu$ is the Kronecker delta. However, this type of tetrad
field gives zero torsion and are of less interest. The non-trivial
tetrad field provides non-zero torsion and yields the construction
of TP theory. It can be written as
\begin{equation}\label{1.1.6}
h_i={h_i}^\mu\partial_\mu,\quad h^j={h^j}_\nu dx^\nu
\end{equation}
satisfying the following properties ${h^i}_\mu{h_j}^\mu=\delta^i_j,~
{h^i}_\mu{h_i}^\nu=\delta_\mu^\nu$. This field establishes metric
tensor as a by product given as follows
\begin{equation}\label{1.1.4}
g_{\mu\nu}=\eta_{ij}{h^i}_\mu{h^j}_\nu.
\end{equation}
The torsion scalar has the following form
\begin{equation}\label{1.3.5}
T={S_\gamma}^{\mu\nu}{T^\gamma}_{\mu\nu}.
\end{equation}
The superpotential tensor ${S_\gamma}^{\mu\nu}$ (anti-symmetric in
its upper indices) is
\begin{equation}\label{1.3.6}
{S_\gamma}^{\mu\nu}=\frac{1}{2}({K^{\mu\nu}}_{\gamma}+\delta^{\mu}_{\gamma}{T^{\theta\nu}}_{\theta}-
\delta^{\nu}_{\gamma}{T^{\theta\mu}}_{\theta}).
\end{equation}
The torsion tensor as follows
\begin{equation}\label{1.2.10}
{T^\gamma}_{\mu\nu}={\widetilde{\Gamma}^\gamma}_{~\nu\mu}-{\widetilde{\Gamma}^\gamma}_{~\mu\nu}
={h_i}^{\gamma}(\partial_{\nu}{h^i}_{\mu}-\partial_{\mu}{h^i}_{\nu}),
\end{equation}
which is antisymmetric in its lower indices, i.e.,
${T^\gamma}_{\mu\nu}=-{T^\gamma}_{\nu\mu}$. The contorsion tensor
can be defined as
\begin{eqnarray}\label{1.2.14}
{K^{\gamma}}_{\mu\nu}=\frac{1}{2}[{{T_\mu}^\gamma}_{\nu}+{{T_\nu}^\gamma}_{\mu}-
{T^\gamma}_{\mu\nu}].
\end{eqnarray}

To formulate a suitable form of the field equations which
establishes the equivalent description (upto equations level) of
these theories, we follow the covariant formalism \cite{S31}.
Incorporating the above equations and some algebraic manipulations,
it follows that
\begin{equation}\label{1117a}
G_{\mu\nu}-\frac{1}{2}g_{\mu\nu}T=-\nabla^{\gamma}S_{\nu\gamma\mu}
-{S^{\sigma\gamma}}_{\mu}K_{\gamma\sigma\nu},
\end{equation}
where $G_{\mu\nu}=R_{\mu\nu}-\frac{1}{2}g_{\mu\nu}R$ is the Einstein
tensor. Finally, we obtain the following field equations for $f(T)$
gravity as
\begin{equation}\label{118a}
f_{T}G_{\mu\nu}+\frac{1}{2}g_{\mu\nu}(f-Tf_T)+E_{\mu\nu}f_{TT}=\kappa^2\mathcal{T}_{\mu\nu},
\end{equation}
where $E_{\mu\nu}={S_{\nu\mu}}^{\gamma}\nabla_{\gamma}T$. It is
noted that Eq.(\ref{118a}) possesses an equivalent structure like
$f(R)$ gravity and reduces to GR for $f(T)=T$. The trace equation is
used to constrain and simplify the field equations. Here, the trace
of the above equation is
\begin{equation}\nonumber
Ef_{TT}-(R+2T)f_T+2f=\kappa^2\mathcal{T},
\end{equation}
with $E={E^\nu}_{\nu}$ and $\mathcal{T}={\mathcal{T}^{\nu}}_{\nu}$.
In terms of effective energy-momentum tensor, the $f(T)$ field
equations can be rewritten as
\begin{equation}\label{120a}
G_{\mu\nu}=\kappa^2\mathcal{T}_{\mu\nu}^{eff}=\kappa^2(\mathcal{T}_{\mu\nu}^{f}+\mathcal{T}_{\mu\nu}^{T}).
\end{equation}
The term
$\mathcal{T}_{\mu\nu}^{f}=\frac{\mathcal{T}_{\mu\nu}^{m}}{f_T}$
corresponds to the matter fluid while using trace equation, torsion
contribution is given by
\begin{equation}\label{122a}
\mathcal{T}_{\mu\nu}^{T}=\frac{1}{\kappa^2f_T}[-E_{\mu\nu}f_{TT}-
\frac{1}{4}g_{\mu\nu}(\mathcal{T}-Ef_{TT}+Rf_{T})].
\end{equation}

\section{Field Equations for Wormhole Construction}

Gravitational theories (like $f(R)$ theory) developed through metric
tensor as well as its dependent quantities are local Lorentz
invariant. Being the basic entity in $f(T)$ gravity, the torsion
tensor (taking tetrad) which induces the TP structure on spacetime
fails to be invariant under local Lorentz transformations
\cite{S31}. This can be seen from
$R=-T-2\nabla^{\gamma}{T^\nu}_{\gamma\nu}$, where $R$ is a covariant
scalar while $T$ as well as $\nabla^{\gamma}{T^\nu}_{\gamma\nu}$
being covariant scalars but not local Lorentz invariant. However,
the later scalar can be neglected inside integral for TP theory and
becomes equivalent to GR.

The non-invariant theories might be sensitive in order to choose
reasonable tetrad which are not uniquely determined by the given
metric $g_{\mu\nu}$. In general, one comes across by a more
complicated connection between the tetrad and metric, particularly
for non-diagonal tetrad with diagonal metric (or even sometimes
diagonal tetrad). Different field equations are developed by taking
different tetrad which successively induce distinct solutions. In an
appropriate limit, some of these solutions lead to GR counterparts
while others fail to provide valid GR counterpart. Thus, the choice
of tetrad is a crucial point in $f(T)$ theory and needs peculiar
attention.

When we deal with spherical coordinates, the diagonal tetrad become
unsuitable as they provide some constraints on $T$ and $f(T)$ model
\cite{S32}. We obtain an unwanted condition $\dot{T}f_{TT}=0$ (or
simply $f_{TT}=0$) which yields torsion scalar to be constant or
$f(T)=c_1+c_2T$ representing TP theory. Thus, the diagonal tetrad do
not represent a useful choice for spherical symmetry. In order to
search for realistic source towards wormhole solutions in $f(T)$
gravity, we assume the following non-diagonal tetrad for static
spherically symmetric wormhole spacetime (\ref{126a})
\begin{center}${h^i}_{\mu}=\left(\begin{array}{cccc}
e^{2\Psi(r)} & 0 & 0 & 0 \\
0 & \frac{1}{\sqrt{1-\frac{b}{r}}}\sin\theta \cos\phi & r\cos\theta\cos\phi & -r\sin\theta\sin\phi \\
0 & \frac{1}{\sqrt{1-\frac{b}{r}}}\sin\theta \sin\phi & r\cos\theta\sin\phi & r\sin\theta\cos\phi \\
0 & \frac{1}{\sqrt{1-\frac{b}{r}}}\cos\theta & -r\sin\theta & 0 \\
\end{array}\right)$.
\end{center}
The torsion scalar turns out to be
\begin{eqnarray}\label{yc3}
T=-\frac{2}{r}\left[2\Psi'\left(\sqrt{1-\frac{b}{r}}-1+\frac{b}{r}\right)-\frac{1}{r}
\left(2\left(1-\sqrt{1-\frac{b}{r}}\right)-\frac{b}{r}\right)
\right].
\end{eqnarray}
We assume anisotropic energy-momentum tensor as
\begin{equation}\label{zc3}
\mathcal{T}^{\mu}_{~\nu}=(\rho+p_t)U^{\mu}U_{\nu}-p_{t}\delta^{\mu}_{\nu}
+(p_r-p_t)V^{\mu}V_{\nu},
\end{equation}
where $p_r$ and $p_t$ are the radial and transverse components of
pressure. The four-velocity of the fluid $U^{\mu}$ and the unit
spacelike vector $V^{\mu}$ satisfy
$U^{\mu}U_{\mu}=1,~V^{\mu}V_{\mu}=-1,~U^{\mu}V_{\mu}=0$. The
corresponding energy-momentum tensor is
\begin{equation*}
\mathcal{T}^{\mu}_{~\nu}=\textmd{diag}(\rho,-p_r,-p_t,-p_t).
\end{equation*}
Equation (\ref{118a}) yields the following field equations
\begin{eqnarray}\label{16c3}
&&\frac{\rho}{f_{_T}}-\frac{1}{r}\left(1-\frac{b}{r}-
\sqrt{1-\frac{b}{r}}\right)T'\frac{f_{_{TT}}}{f_{_T}}-\frac{J}{f_{_T}}=\frac{b'}{r^2},\\\label{17c3}
&&\frac{p_r}{f_{_T}}+\frac{J}{f_{_T}}=\frac{1}{r^2}\left[2r\Psi'\left(1-\frac{b}{r}\right)
-\frac{b}{r}\right],\\\nonumber
&&\frac{p_t}{f_{_T}}+\frac{1}{2}\left\{\Psi'\left(1-\frac{b}{r}\right)-\frac{1}{r}\left(\sqrt{1-\frac{b}{r}}-
1+\frac{b}{r}\right)\right\}
T'\frac{f_{_{TT}}}{f_{_T}}+\frac{J}{f_{_T}}=\frac{1}{2r^2}\left[2r\Psi'\right.\\\label{18c3}&&-\left.\Psi'b-b'+
\frac{b}{r}+2\Psi''r^2-2\Psi''rb+2\Psi'^2r^2-2\Psi'^2rb-\Psi'rb'\right],
\end{eqnarray}
where prime refers derivative with respect to $r$ and $J(r)$ is
given by
\begin{equation}\nonumber
J(r)=\frac{1}{4}(\mathcal{T}-Ef_{_{TT}}+Rf_{_T}).
\end{equation}

Taking effective energy density and pressure from Eqs.(\ref{16c3})
and (\ref{17c3}), NEC yields
\begin{equation}\label{31c3}
\rho^{eff}+p_r^{eff}=\frac{b'r-b}{r^3}+\frac{2}{r}\left(1-\frac{b}{r}\right)\Psi'.
\end{equation}
Due to flaring out condition, we obtain $b>b'r$ leading to the
violation of NEC with $\Psi'<0$, i.e., $\rho^{eff}+p_r^{eff}\leq0$
which implies that the effective energy-momentum tensor is
responsible for the necessary violation of energy conditions to
support wormhole geometry. Thus, it may impose condition on the
usual matter to satisfy the energy conditions in this scenario and
establish some physically acceptable solutions.

To be traversable wormhole solution, the magnitude of its redshift
function must be finite. Also, upto equation level for constant
value of $\Psi$ other than zero, we note that only $\Psi'$ is used
for which $\Psi=$constant gives same scenario. For the sake of
simplicity, setting $\Psi=0$ in Eqs.(\ref{16c3})-(\ref{18c3}), the
field equations can be written as
\begin{eqnarray}\label{20c3}
\rho &=&\frac{b'}{r^2}f_{_T}+\frac{1}{r}\left(1-\frac{b}{r}-
\sqrt{1-\frac{b}{r}}\right)T'f_{_{TT}},\\\label{21c3} p_r
&=&-\frac{b}{r^3}f_T,\\\label{22c3}
p_t&=&\frac{1}{2r}\left(\sqrt{1-\frac{b}{r}}-
1+\frac{b}{r}\right)T'f_{_{TT}}-
\frac{1}{2r^2}\left(b'-\frac{b}{r}\right)f_{_T},
\end{eqnarray}
where the torsion scalar becomes
\begin{equation}\label{fgc3}
T=\frac{2}{r^2}\left[2-\frac{b}{r}-2\sqrt{1-\frac{b}{r}}\right].
\end{equation}
It is noted that Eq.(\ref{31c3}) gives the violation of NEC in terms
of flaring out condition.

\section{Wormhole Solutions}

Non-commutative geometry is the intrinsic characteristic of
spacetime and plays an effective role in several areas. To formulate
NC form of GR, the coordinate coherent state approach is widely
used. The NC geometry is used to eliminate the divergencies that
appear in GR by replacing the point-like structures with smeared
objects. Taking Lorentzian distribution, the energy density of the
particle-like static spherically symmetric gravitational source
having mass $M$ takes the following form \cite{24c3}
\begin{equation}\label{ac3}
\rho_{_{NCL}}=\frac{M\sqrt{\phi}}{\pi^2(r^2+\phi)^2},
\end{equation}
where $\phi$ is the NC parameter. Taking into account correspondence
between $\rho$ and $\rho_{_{NCL}}$ using Eqs.(\ref{20c3}) and
(\ref{ac3}), we obtain the following differential equation
\begin{eqnarray}\label{15'}
\frac{b'}{r^2}f_{_T}+\frac{1}{r}\left(1-\frac{b}{r}-
\sqrt{1-\frac{b}{r}}\right)T'f_{_{TT}}=
\frac{M\sqrt{\phi}}{\pi^2(r^2+\phi)^2},
\end{eqnarray}
which contains two unknown functions, $b$ and $f$. Thus, we have to
choose one of these function and carried out steps for the other
one. We discuss the NCL wormhole solutions in $f(T)$ gravity for
non-diagonal tetrad and investigate the behavior of energy
conditions for specific $f(T)$ and shape functions.

\subsection{For Power-law $f(T)$ Model}

We assume model in power-law form of torsion scalar which is the
generalization of GR and analogy to $f(R)$ model like $f(R)=R+\delta
R^2$ taken to discuss the wormhole solutions. The $f(T)$ model is
\begin{equation}\label{a}
f(T)=T+\alpha T^2,
\end{equation}
This model has contributed as the most viable model due to its
simple form and we may directly compare our results with GR. We
discuss wormhole solutions for the following two cases:

\subsection*{Case I: $\alpha=0$}

We consider $\alpha=0$ in model (\ref{a}) which leads the whole
scenario in teleparallel gravity. In this case, Eq.(\ref{15'})
becomes
\begin{eqnarray}\label{15''}
\frac{b'}{r^2}= \frac{M\sqrt{\phi}}{\pi^2(r^2+\phi)^2},
\end{eqnarray}
yields the solution
\begin{equation}\label{38}
b(r)=\frac{M}{2\pi^2}\left(\frac{r\sqrt{\phi}}{r^2+\phi}+\tan^{-1}(\frac{r}{\sqrt{\phi}})\right)+c,
\end{equation}
where $c$ is an arbitrary constant.
\begin{figure} \centering
\epsfig{file=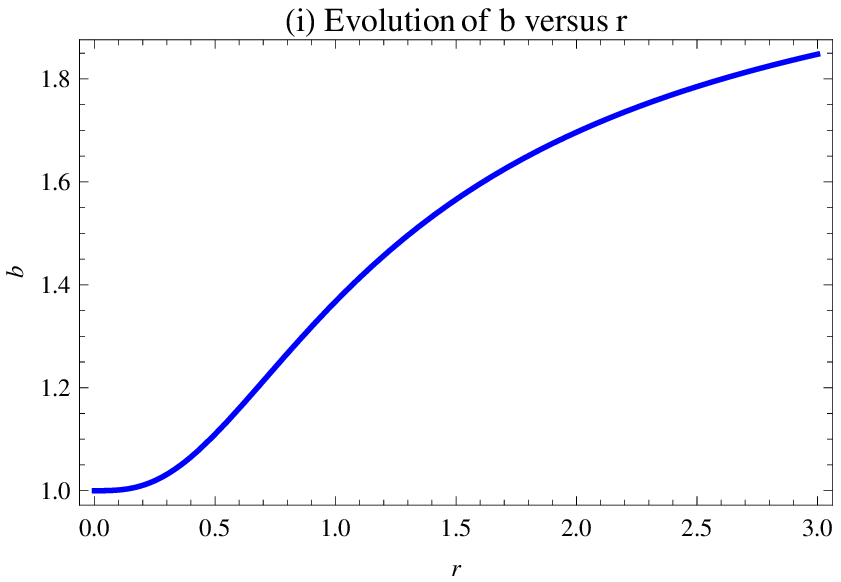,width=.45\linewidth}\epsfig{file=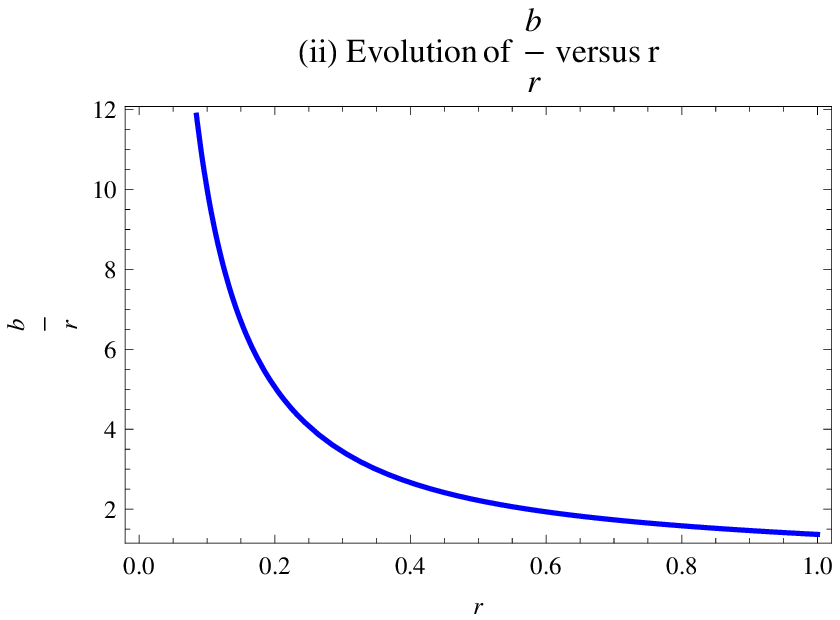,width=.45\linewidth}
\epsfig{file=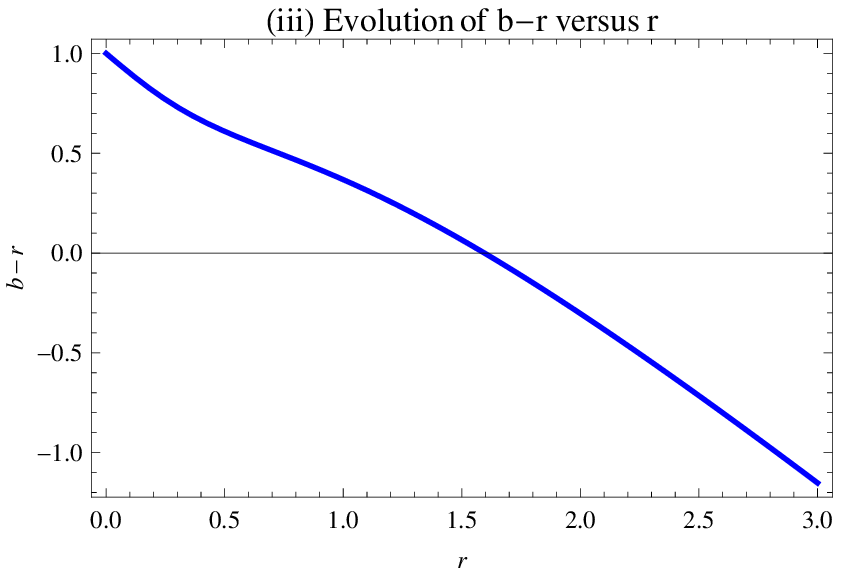,width=.45\linewidth}\epsfig{file=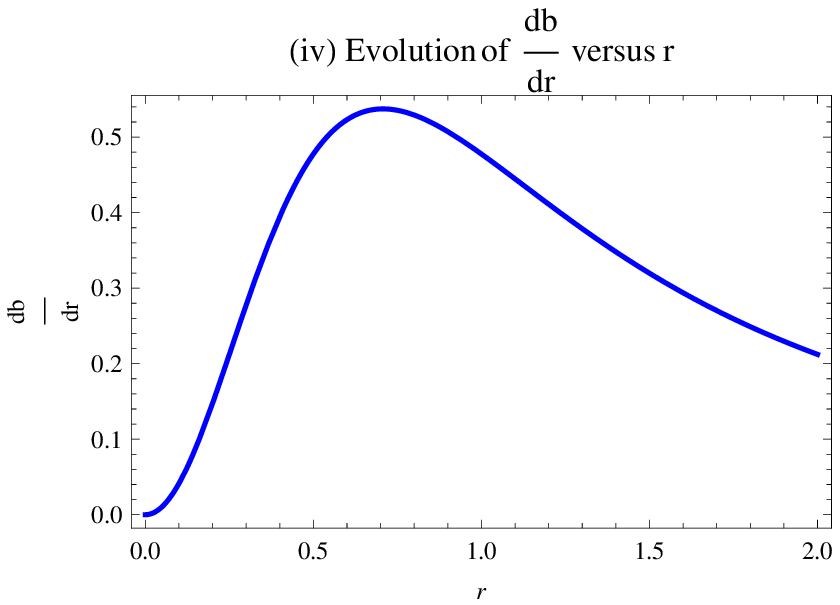,width=.45\linewidth}\caption{Plots
of shape function in teleparallel case: (i) $b$ versus $r$, (ii)
$\frac{b}{r}$ versus $r$, (iii) $b-r$ versus $r$, (iv)
$\frac{db}{dr}$ versus $r$.}
\end{figure}

In order to examine the geometry of wormhole, we draw the shape
function taking different conditions in \textbf{Figure 1}. We choose
arbitrarily the values of model parameters, such as,
$\phi=0.5,~M=15$ and $c=1$. \textbf{Figure 1(i)} represents the
evolution of shape function in increasing manner versus $r$. For
asymptomatically flat condition, we draw $\frac{b}{r}$ with respect
to increasing $r$. It can be seen from plot \textbf{(ii)} that
$\frac{b}{r}$ approaches towards $0$ as $r\rightarrow 0$. This
implies that asymptotically flat condition for wormhole construction
is satisfied. In plot \textbf{(iii)}, we draw $b-r$ versus $r$ to
find out throat radius. It is noted that throat radius is that
minimum value for which $b-r$ cuts the $r$-axis. In this case, the
throat radius is obtained as $r_0=1.6$. This plot satisfies the
condition $1-\frac{b}{r}>0$ for $r>r_0$. In Figure \textbf{1(iv)},
we plot the first derivative of $b(r)$ with respect to $r$ to check
the validity of condition $b'(r_0)<1$ which shows that the
corresponding condition is satisfied. Thus, the shape function
satisfies the required structure of the wormhole.
\begin{figure} \centering
\epsfig{file=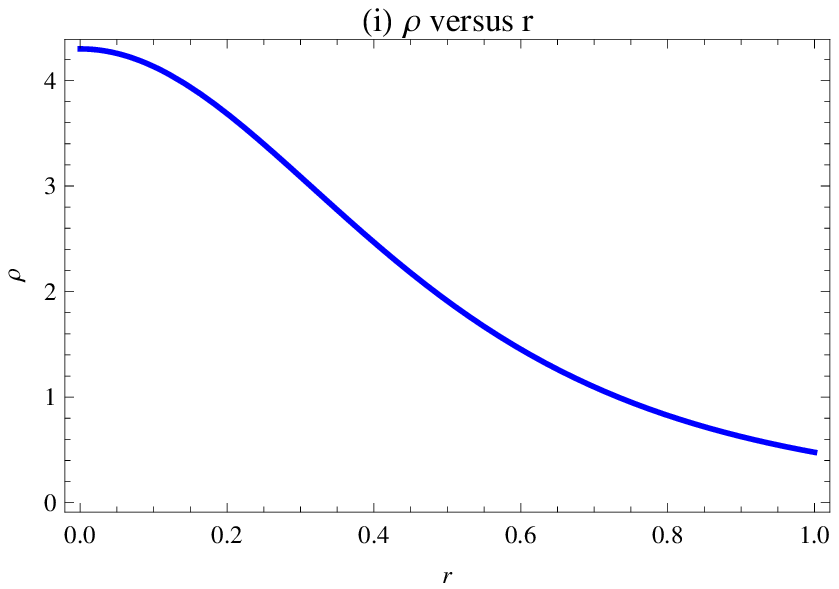,width=.45\linewidth}\epsfig{file=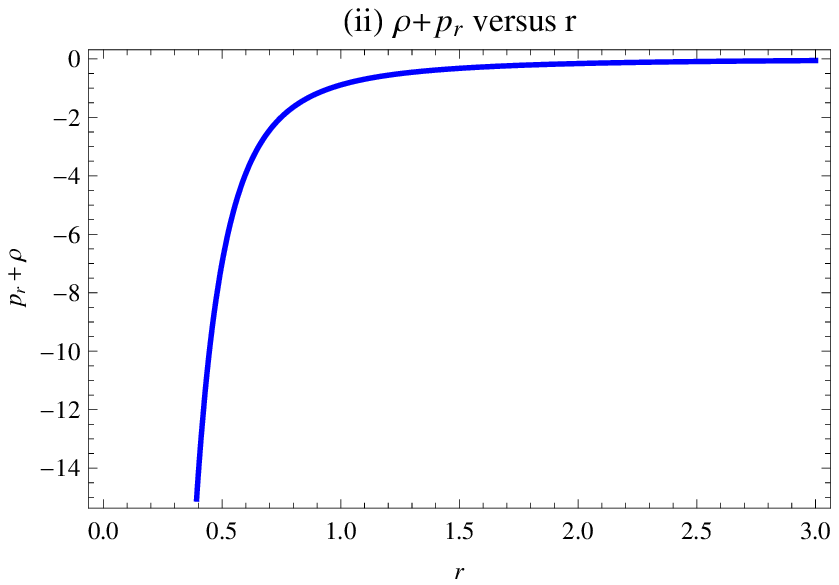,width=.45\linewidth}
\epsfig{file=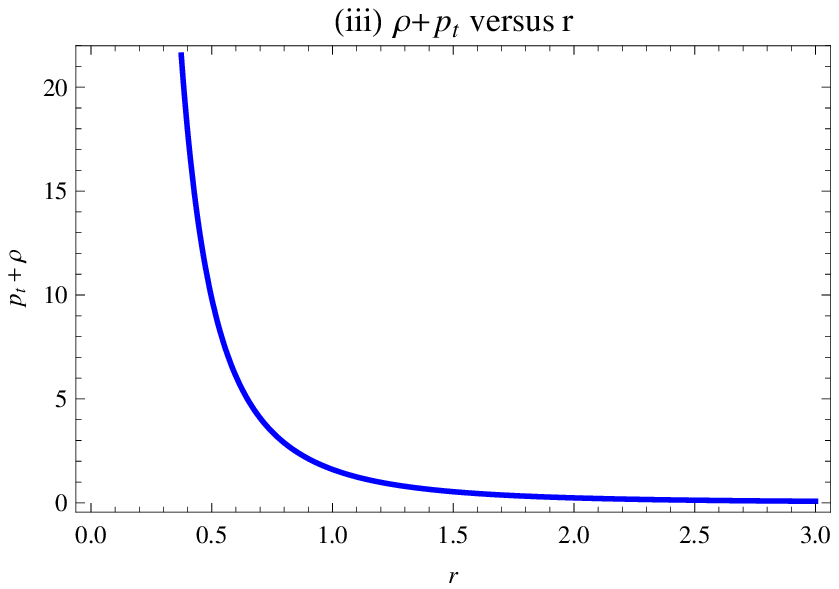,width=.45\linewidth}\caption{Plots of WEC in
teleparallel case: (i) $\rho$ versus $r$, (ii) $\rho+p_r$ versus
$r$, (iii) $\rho+p_t$ versus $r$.}
\end{figure}

Equations (\ref{20c3})-(\ref{22c3}) are given as follows
\begin{eqnarray}\label{20c3'}
\rho &=&\frac{M\sqrt{\phi}}{\pi^2(r^2+\phi)^2},\\\label{21c3'} p_r
&=&-\frac{1}{r^3}\left[\frac{M}{2\pi^2}\left(\frac{r\sqrt{\phi}}{r^2+\phi}+
\tan^{-1}(\frac{r}{\sqrt{\phi}})\right)+c\right],\\\label{22c3'}
p_t&=&-\frac{1}{2r^2}\left[\frac{M\sqrt{\phi}}{\pi^2(r^2+\phi)^2}-\frac{1}{r}\left\{\frac{M}{2\pi^2}
\left(\frac{r\sqrt{\phi}}{r^2+\phi}+
\tan^{-1}(\frac{r}{\sqrt{\phi}})\right)+c\right\}\right].
\end{eqnarray}
The behavior of WEC ($\rho,~\rho+p_r$ and $\rho+p_t$) versus $r$ is
shown in Figure \textbf{2}. The curves of $\rho,~\rho+p_t$ represent
positively decreasing behavior for increasing $r$ while $\rho+p_r$
indicates negative behavior which shows the violation of WEC. Thus
no physically acceptable wormhole solution is obtained in
teleparallel case. These results are compatible with the results of
\cite{fr}.

\subsection*{Case II: $\alpha\neq0$}
\begin{figure} \centering
\epsfig{file=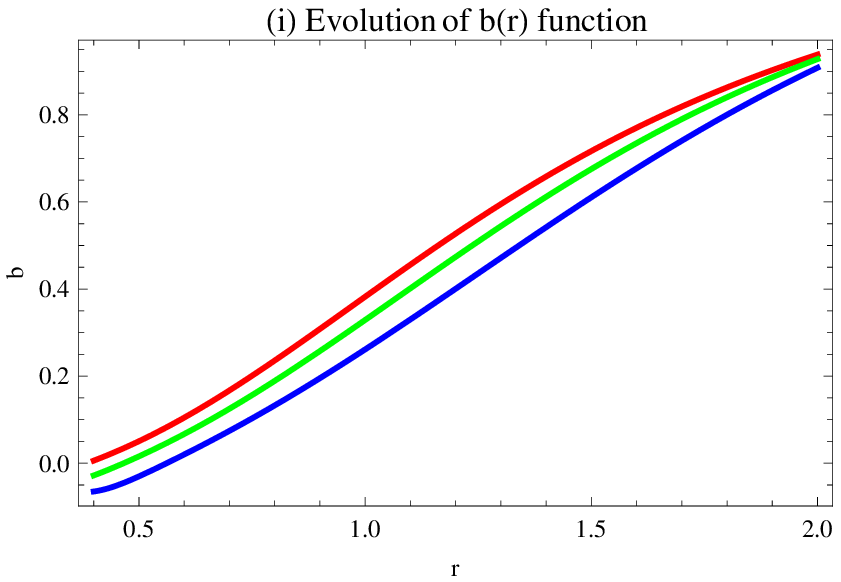,width=.45\linewidth}\epsfig{file=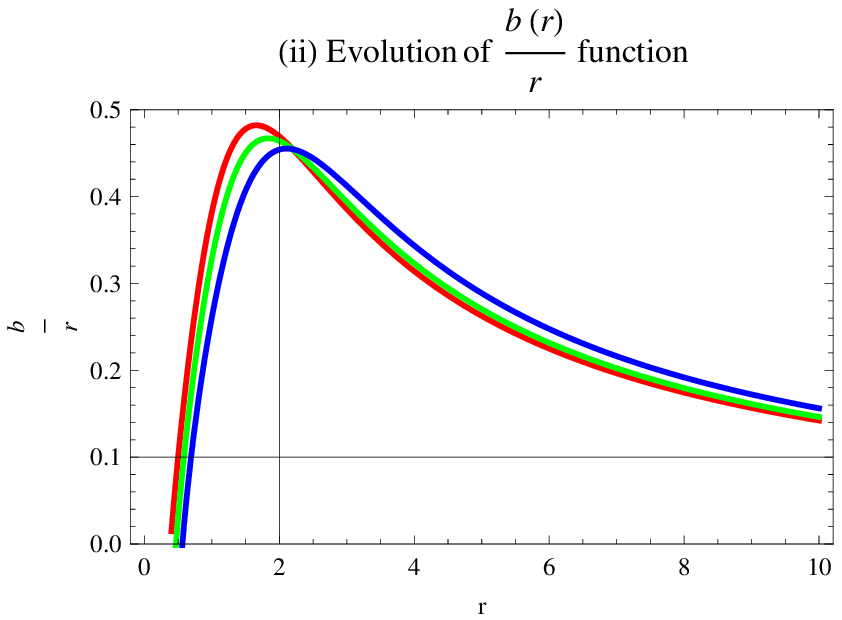,width=.45\linewidth}
\epsfig{file=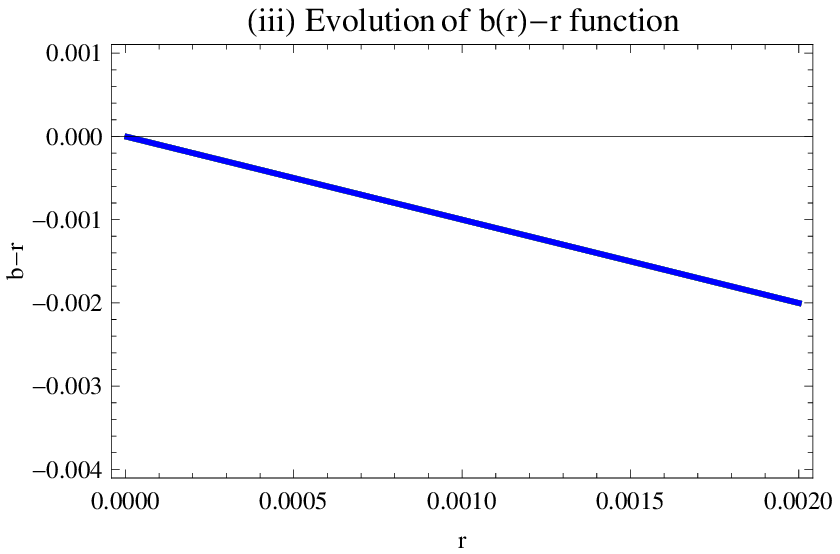,width=.45\linewidth}\epsfig{file=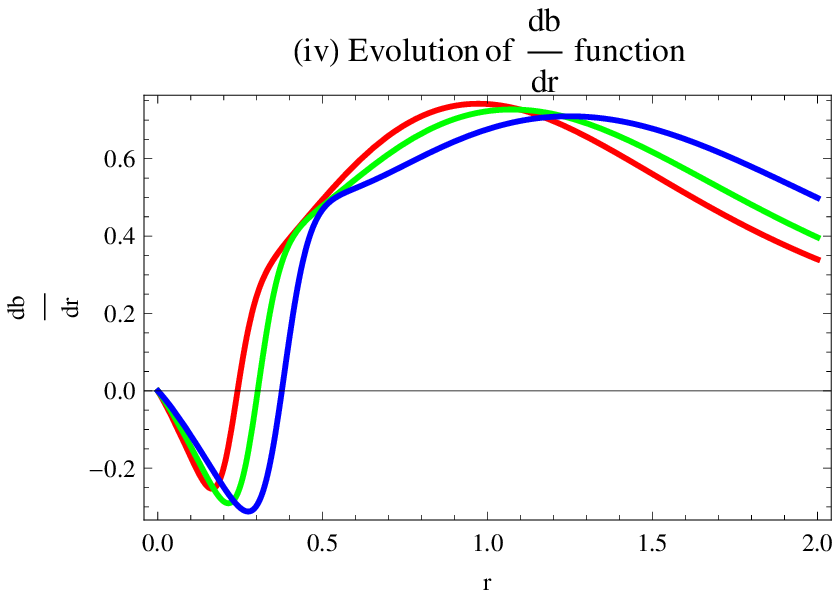,width=.45\linewidth}\caption{Plots
of shape function in $f(T)$ case: red curve for $\alpha=-2$, green
curve for $\alpha=-3$, blue curve for $\alpha=-5$ (i) $b$ versus
$r$, (ii) $\frac{b}{r}$ versus $r$, (iii) $b-r$ versus $r$, (iv)
$\frac{db}{dr}$ versus $r$.}
\end{figure}
The case $\alpha\neq0$ leads to the extended teleparallel gravity.
Inserting Eq.(\ref{a}) in (\ref{15'}), we obtain the following
differential equation
\begin{eqnarray}\nonumber
\frac{r^2M\sqrt{\phi}}{\pi^2(r^2+\phi)^2}&=&b'\left\{1+\frac{4\alpha}{r^2}\left(2-\frac{b}{r}-
2\sqrt{1-\frac{b}{r}}\right)\right\}-\frac{4\alpha}{r^2}\left(1-\frac{b}{r}-\sqrt{1-\frac{b}{r}}
\right)\\\label{16'}&\times&\left[4\left(1-\sqrt{1-\frac{b}{r}}\right)
-\frac{2b}{r}+\frac{b'r-b}{r}\left(1-\frac{1}{\sqrt{1-\frac{b}{r}}}\right)\right].
\end{eqnarray}
We check the behavior of shape function and flaring out conditions
numerically through graphs by taking same values of parameters along
with three different values of $\alpha$ such as $\alpha=-2,~-3,~-5$
and initial value as $f(2.2)=1$. Figure \textbf{3(i)} represents
increasing behavior of the shape function versus $r$. In the right
graph \textbf{(ii)}, $\frac{b}{r}$ versus $r$ shows that
$\frac{b}{r}$ approaches to zero as we increase $r$ which represents
that asymptomatically flatness is obtained. To locate throat of the
wormhole, we plot $b(r)-r$ versus $r$ as shown in the \textbf{(iii)}
plot of Figure \textbf{3}. In this plot, the throat is located at
very small values of $r$. The first derivative of shape function is
also plotted versus $r$ as shown in \textbf{(iv)} plot which
indicates that $\frac{db}{dr}$ at $r_0$ satisfies the condition,
$b'(r_0)<1$. Thus similar to teleparallel case, shape function
satisfies the conditions of wormhole geometry.

To check the behavior of WEC for power-law model, the expressions of
matter content using Eqs.(\ref{20c3})-(\ref{22c3}) are given by
\begin{eqnarray}\nonumber
\rho&=&\frac{b'}{r^2}\left\{1+\frac{4\alpha}{r^2}\left(2-\frac{b}{r}-
2\sqrt{1-\frac{b}{r}}\right)\right\}-\frac{4\alpha}{r^2}\left(1-\frac{b}{r}-\sqrt{1-\frac{b}{r}}
\right)\\\label{3c3}&\times&\left[4\left(1-\sqrt{1-\frac{b}{r}}\right)
-\frac{2b}{r}+\frac{b'r-b}{r}\left(1-\frac{1}{\sqrt{1-\frac{b}{r}}}\right)\right],\\\label{4c3}
p_r&=&
-\frac{b}{r^3}\left\{1+\frac{4\alpha}{r^2}\left(2-\frac{b}{r}-
2\sqrt{1-\frac{b}{r}}\right)\right\},\\\nonumber
p_t&=&-\frac{1}{2r^2}\left(b'-\frac{b}{r}\right)\left\{1+\frac{4\alpha}{r^2}\left(2-\frac{b}{r}-
2\sqrt{1-\frac{b}{r}}\right)\right\}-\frac{2\alpha}{r^4}\left(1-\frac{b}{r}\right.\\\label{5c3}
&-&\left.\sqrt{1-\frac{b}{r}}
\right)\left[4\left(1-\sqrt{1-\frac{b}{r}}\right)
-\frac{2b}{r}+\frac{b'r-b}{r}\left(1-\frac{1}{\sqrt{1-\frac{b}{r}}}\right)\right].
\end{eqnarray}
Figure \textbf{4} represents graph of WEC expressions versus $r$
which show that $\rho$ and $\rho+p_t$ show decreasing behavior but
remain positive. For $\alpha=-2,~\rho$ indicates negative value at
$r\leq0.52$. The behavior of $\rho+p_r$ is negative but there exist
some part of the curves in positive panel of plot. Thus there exist
possibility to have micro or tiny wormhole.
\begin{figure} \centering
\epsfig{file=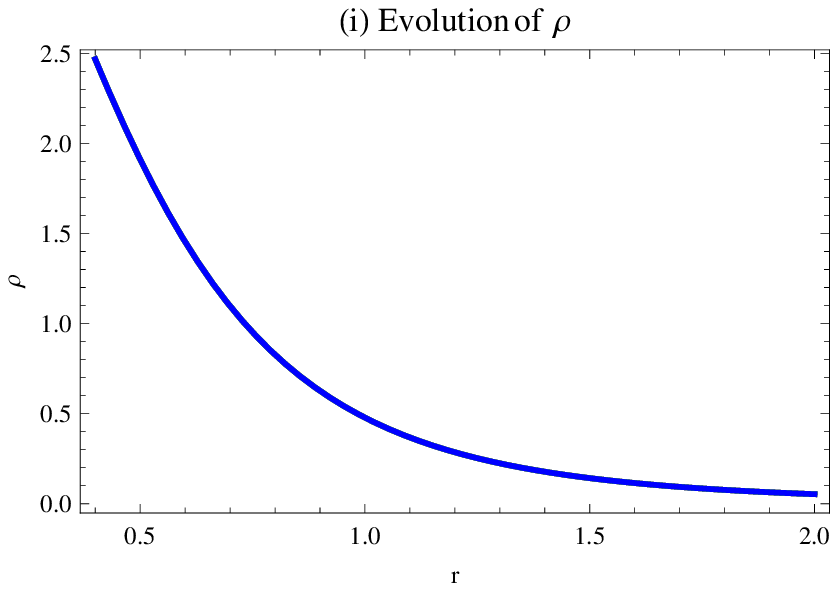,width=.45\linewidth}\epsfig{file=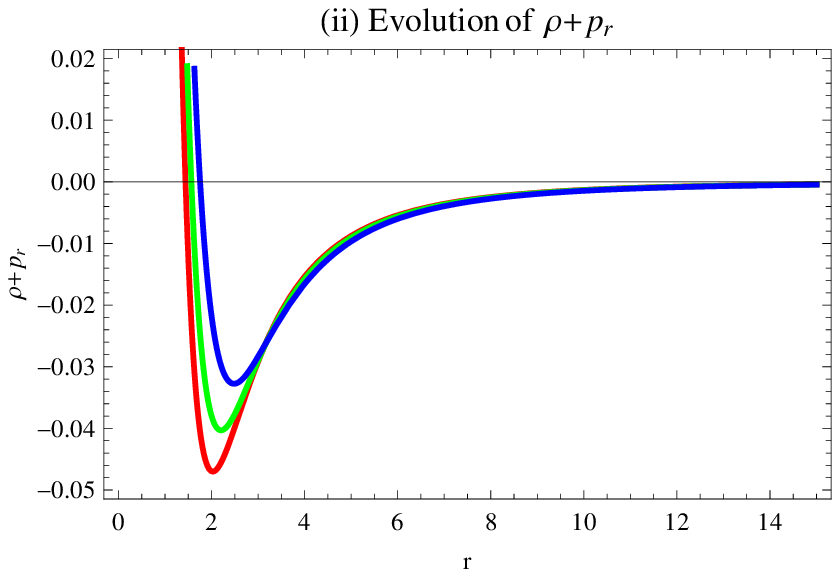,width=.45\linewidth}
\epsfig{file=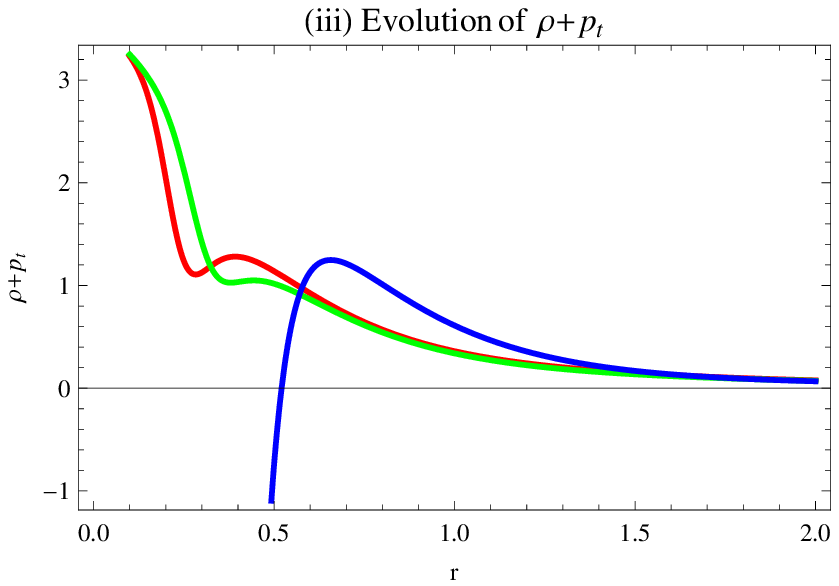,width=.45\linewidth}\caption{Plots of WEC in
$f(T)$ case: red curve for $\alpha=-2$, green curve for $\alpha=-3$,
blue curve for $\alpha=-5$ (i) $\rho$ versus $r$, (ii) $\rho+p_r$
versus $r$, (iii) $\rho+p_t$ versus $r$.}
\end{figure}

\subsection{For Power-law $b(r)$ Model}

Here, we examine the wormhole solution by considering a specific
shape function in terms of power-law form and construct $f(T)$
function in the NCL background. We take the following particular
shape function \cite{22c3,29c3}
\begin{equation}\label{44c33}
b(r)=r_0\left(\frac{r}{r_0}\right)^{\gamma},
\end{equation}
where $\gamma$ is any constant. To meet the wormhole geometry, it
can be seen that $b'(r_0)<1$ implies that $b'(r_0)=\gamma<1$ and
$b(r_0)=r_0$ holds. The asymptotically flat spacetime is also
obtained for this shape function, i.e.,
$\frac{b(r)}{r}=r_0^{1-\gamma}r^{\gamma-1}\rightarrow 0$ as
$r\rightarrow\infty$. Substituting the values of $\rho_{_{NCL}}$ and
$b(r)$ from Eqs.(\ref{ac3}) and (\ref{44c33}) in (\ref{20c3}), we
obtain the following differential equation
\begin{eqnarray}\nonumber
&&\frac{1}{r}\left[1-\left(\frac{r}{r_0}\right)^{\gamma-1}-\sqrt{1-\left(\frac{r}{r_0}\right)^{\gamma-1}}\right]
\frac{f''}{T'}+\frac{\gamma}{r^2}\left(\frac{r}{r_0}\right)^{\gamma-1}\frac{f'}{T'}
-\frac{1}{r}\\\label{45c3}&&
\times\left[1-\left(\frac{r}{r_0}\right)^{\gamma-1}-\sqrt{1-\left(\frac{r}{r_0}\right)^{\gamma-1}}\right]
\frac{T'' f'}{T'^2}=\frac{M\sqrt{\phi}}{\pi^2(r^2+\phi)^2},
\end{eqnarray}
where
\begin{eqnarray*}
T&=&\frac{2}{r^2}\left[2-\left(\frac{r}{r_0}\right)^{\gamma-1}-2
\sqrt{1-\left(\frac{r}{r_0}\right)^{\gamma-1}}\right],\\
T'&=&-\frac{2}{r^3}\left[4\left(1-\sqrt{1-\left(\frac{r}{r_0}\right)^{\gamma-1}}\right)
-2\left(\frac{r}{r_0}\right)^{\gamma-1}+(\gamma-1)\right.\\&\times&\left.\left(\frac{r}{r_0}\right)^{\gamma-1}
\left(1-\frac{1}{\sqrt{1-\left(\frac{r}{r_0}\right)^{\gamma-1}}}\right)\right],
\\\nonumber
T''&=&\frac{12}{r^4}\left[2\left(1-\sqrt{1-\left(\frac{r}{r_0}\right)^{\gamma-1}}\right)
-\left(\frac{r}{r_0}\right)^{\gamma-1}-\frac{(\gamma-1)(\gamma-6)}{6}\right.
\\&\times&\left. \left(\frac{r}{r_0}\right)^{\gamma-1}
\left(1-\frac{1}{\sqrt{1-\left(\frac{r}{r_0}\right)^{\gamma-1}}}\right)+\frac{(\gamma-1)^2}{12}
\left(\frac{r}{r_0}\right)^{2(\gamma-1)}\right. \\&\times&\left.
\left(1-\left(\frac{r}{r_0}\right)^{\gamma-1}\right)^{-\frac{3}{2}}\right].
\end{eqnarray*}
We evaluate $f(T)$ function numerically and draw its plot as well as
WEC versus $T$ and $r$ respectively, as shown in Figure \textbf{5}.
Keeping the same values of NCL parameters along with $\gamma=0.5$
for three values of throat radius $r_0=0.93,~0.95,~0.99$. The plot
\textbf{(i)} indicates the positively decreasing behavior of $f$.
\begin{figure} \centering
\epsfig{file=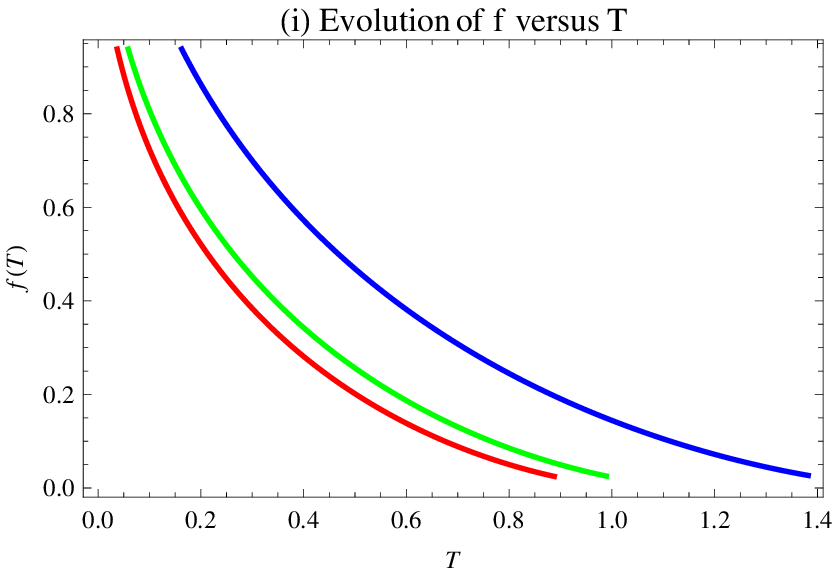,width=.45\linewidth}\epsfig{file=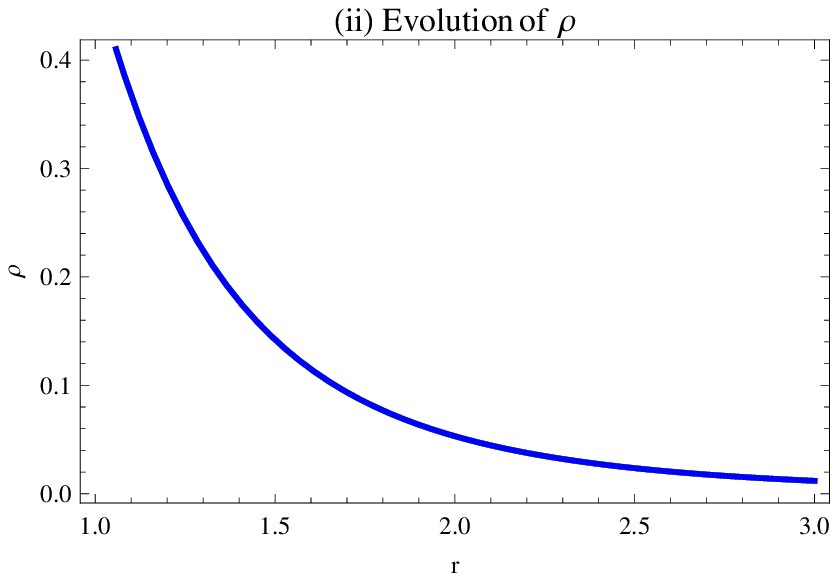,width=.45\linewidth}
\epsfig{file=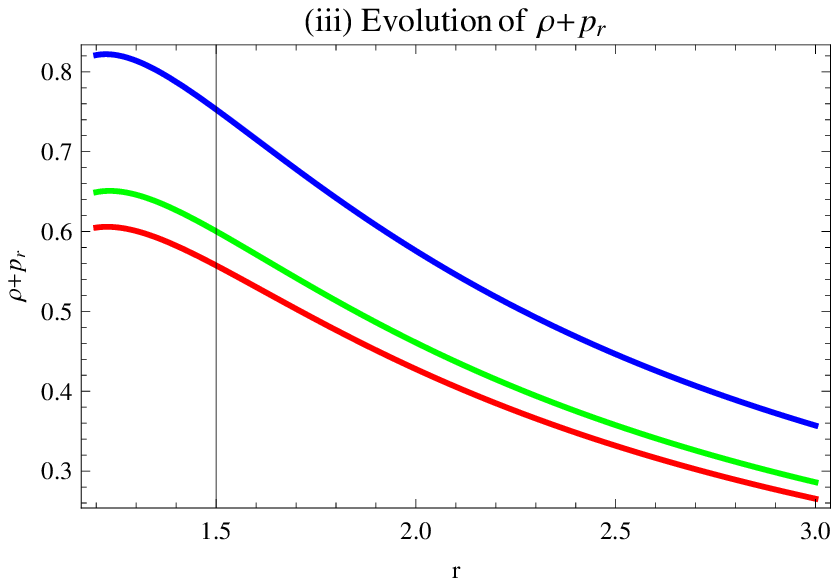,width=.45\linewidth}\epsfig{file=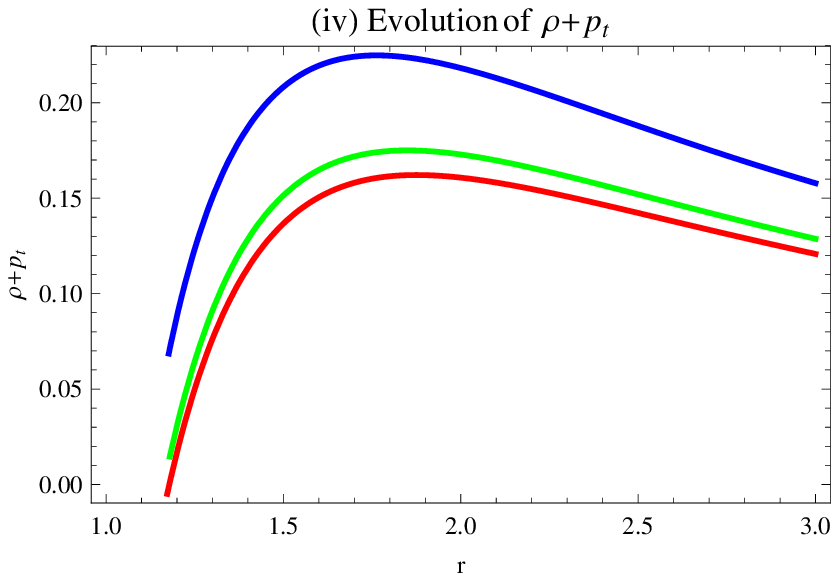,width=.45\linewidth}\caption{Plots
of $f(T)$ and WEC: red curve for $r_0=0.93$, green curve for
$r_0=0.95$, blue curve for $r_0=0.99$ (i) $f(T)$ versus $T$, (ii)
$\rho$ versus $r$, (iii) $\rho+p_r$ versus $r$, (iv) $\rho+p_t$
versus $r$.}
\end{figure}

The expressions for WEC become
\begin{eqnarray}\nonumber
\rho&=&\frac{\gamma}{r^2}\left(\frac{r}{r_0}\right)^{\gamma-1}\frac{f'}{T'}+
\frac{1}{r}\left(1-\left(\frac{r}{r_0}\right)^{\gamma-1}-\sqrt{1-\left(\frac{r}{r_0}\right)^{\gamma-1}}\right)
\\\label{46c3}&\times&\left(\frac{f''}{T'}-\frac{T''f'}{T'^2}\right),\\\nonumber
\rho+p_r&=&\frac{\gamma-1}{r^2}\left(\frac{r}{r_0}\right)^{\gamma-1}\frac{f'}{T'}+
\frac{1}{r}\left(1-\left(\frac{r}{r_0}\right)^{\gamma-1}-\sqrt{1-\left(\frac{r}{r_0}\right)^{\gamma-1}}\right)
\\\label{46c33}&\times&\left(\frac{f''}{T'}-\frac{T''f'}{T'^2}\right),\\\nonumber
\rho+p_t&=&\frac{\gamma
+1}{2r^2}\left(\frac{r}{r_0}\right)^{\gamma-1}\frac{f'}{T'}+
\frac{1}{2r}\left(1-\left(\frac{r}{r_0}\right)^{\gamma-1}-\sqrt{1-\left(\frac{r}{r_0}\right)^{\gamma
-1}}\right)
\\\label{47c3}&\times&\left(\frac{f''}{T'}-\frac{T''f'}{T'^2}\right).
\end{eqnarray}
Figure \textbf{5(ii)-(iv)} shows the plots of these expressions
versus $r$. This represents that $\rho,~\rho+p_r,~\rho+p_t$ indicate
positive behavior of these expressions. Thus, physically acceptable
wormhole solutions are obtained for the specific shape function
where the basic role is played by effective energy-momentum tensor.

\section{Equilibrium Conditions}

To discuss the equilibrium configuration of the wormhole solutions,
we consider the generalized Tolman-Oppenheimer-Volkov equation
\cite{S48,S49}
\begin{equation}\label{25c33}
\frac{dp_r}{dr}+\frac{\sigma'}{2}(\rho+p_r)+\frac{2}{r}(p_r-p_t)=0,
\end{equation}
for the metric
$ds^2=\textmd{diag}(e^{\sigma(r)},-e^{\mu(r)},-r^2,-r^2\sin^2\theta)$.
Ponce de Le$\acute{o}$n suggested this equation for anisotropic mass
distribution as follows
\begin{equation}\label{26c33}
\frac{2}{r}(p_t-p_r)-\frac{e^{\frac{\mu-\sigma}{2}}\mathcal{M}_{_{eff}}}{r^2}(\rho+p_r)-\frac{dp_r}{dr}=0.
\end{equation}
Here
$\mathcal{M}_{_{eff}}=\frac{1}{2}r^2e^{\frac{\sigma-\mu}{2}}\sigma'$
is the effective gravitational mass which is measured from throat to
some arbitrary radius $r$. This equation indicates the equilibrium
configuration for the wormhole solutions by taking gravitational,
hydrostatic as well as anisotropic force due to anisotropic matter
distribution. Using Eq.(\ref{26c33}), these forces are defined
respectively as
\begin{equation}\nonumber
\mathcal{F}_{gf}=-\frac{\sigma'(\rho+p_r)}{2},\quad
\mathcal{F}_{hf}=-\frac{dp_r}{dr}, \quad
\mathcal{F}_{af}=\frac{2(p_t-p_r)}{r}.
\end{equation}
For the wormhole solutions to be in equilibrium, it is required that
\begin{equation}\label{28c33}
\mathcal{F}_{gf}+\mathcal{F}_{hf}+\mathcal{F}_{af}=0.
\end{equation}
It is noted here that $\sigma$ represents the gravitational redshift
which is taken as constant, i.e., $\sigma=2\Psi$ leads to
$\sigma'=0$ for constant $\Psi$. This yields $\mathcal{F}_{gf}$
becomes zero and we are left with hydrostatic and anisotropic forces
with corresponding equilibrium condition
$\mathcal{F}_{hf}+\mathcal{F}_{af}=0$. For teleparallel, specific
$f(T)$ and $b(r)$ cases, $\mathcal{F}_{hf}$ and $\mathcal{F}_{af}$
takes the following form respectively
\begin{eqnarray*}
\mathcal{F}_{hf}&=&-\frac{3}{r^4}\left[\frac{M}{2\pi^2}\left(\frac{r\sqrt{\phi}}{r^2+\phi}
+\tan^{-1}\left(\frac{r}{\sqrt{\phi}}\right)\right)+c\right]+\frac{1}{r^3}\left[
\frac{M\sqrt{\phi}(5r^3+3\phi)}{2\pi^2r^3(r^2+\phi)^2}\right.\\&-&\left.\frac{3c}{r^4}-\frac{3M}{2\pi
r^4}\tan^{-1}\left(\frac{r}{\sqrt{\phi}}\right)\right],\\
\mathcal{F}_{af}&=&-\frac{1}{r^3}\left[\frac{M\sqrt{\phi}}{\pi^2(r^2+\phi)^2}-\frac{3}{r}\left\{\frac{M}{2\pi^2}
\left(\frac{r\sqrt{\phi}}{r^2+\phi}+
\tan^{-1}\left(\frac{r}{\sqrt{\phi}}\right)\right)+c\right\}\right],\\
\mathcal{F}_{hf}&=&\left(\frac{b'}{r^3}-\frac{3b}{r^4}\right)\left\{1+\frac{4\alpha}{r^2}\left(2-\frac{b}{r}-
2\sqrt{1-\frac{b}{r}}\right)\right\}+\frac{b}{r^6}\left[-8\alpha\right.\\&\times&\left.\left(2-\frac{b}{r}-
2\sqrt{1-\frac{b}{r}}\right)-\frac{4\alpha(b'r-b)}{r}\left(1-\frac{1}{\sqrt{1-\frac{b}{r}}}\right)\right],\\
\mathcal{F}_{af}&=&-\frac{1}{r^3}\left[\left(b'-\frac{3b}{r}\right)\left\{1+\frac{4\alpha}{r^2}\left(2-\frac{b}{r}-
2\sqrt{1-\frac{b}{r}}\right)\right\}+\frac{4\alpha}{r}\left(\frac{b}{r}-1\right.\right.\\&+&\left.\left.\sqrt{1-\frac{b}{r}}
\right)\left[4\left(1-\sqrt{1-\frac{b}{r}}\right)
-\frac{2b}{r}+\frac{b'r-b}{r}\left(1-\frac{1}{\sqrt{1-\frac{b}{r}}}\right)\right]\right],\\
\mathcal{F}_{hf}&=&\frac{\gamma-3}{r^3}\left(\frac{r}{r_0}\right)^{\gamma-1}\frac{f'}{T'}+
\frac{1}{r^2}\left(\frac{r}{r_0}\right)^{\gamma-1}\left(\frac{f''}{T'}-\frac{T''f'}{T'^2}\right),\\
\mathcal{F}_{af}&=&\frac{3-\gamma}{r^3}\left(\frac{r}{r_0}\right)^{\gamma-1}\frac{f'}{T'}+
\frac{1}{r^2}\left(\left(\frac{r}{r_0}\right)^{\gamma-1}-1+\sqrt{1-\left(\frac{r}{r_0}\right)^{\gamma
-1}}\right)\\&\times&\left(\frac{f''}{T'}-\frac{T''f'}{T'^2}\right).
\end{eqnarray*}
\begin{figure} \centering
\epsfig{file=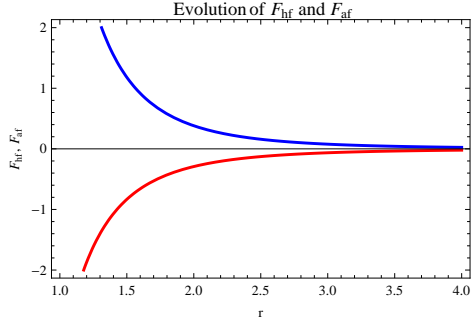,width=.45\linewidth}\caption{Plot of
$\mathcal{F}_{hf}$ and $\mathcal{F}_{af}$ in teleparallel case
versus $r$: red curve represents $\mathcal{F}_{hf}$ and blue curve
represents $\mathcal{F}_{af}$.}
\end{figure}
\begin{figure} \centering
\epsfig{file=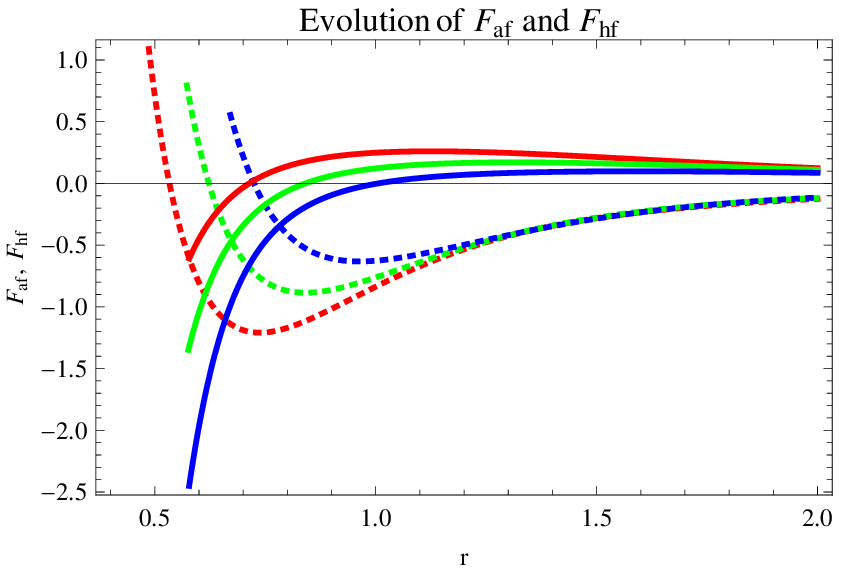,width=.45\linewidth}\caption{Plot of
$\mathcal{F}_{hf}$ (dotted) and $\mathcal{F}_{af}$ (solid) in
specific $f(T)$ case versus $r$.}
\end{figure}
\begin{figure} \centering
\epsfig{file=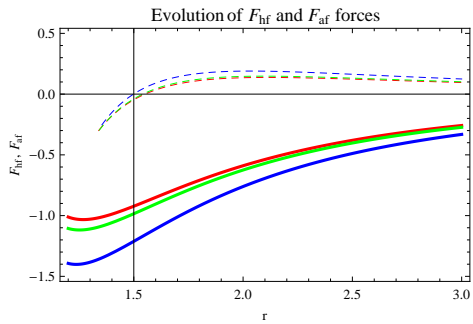,width=.45\linewidth}\caption{Plot of
$\mathcal{F}_{hf}$ (dotted) and $\mathcal{F}_{af}$ (solid) in
specific $b(r)$ case versus $r$.}
\end{figure}

We plot these equations for the obtained wormhole solutions as shown
in \textbf{Figures 6-8} respectively. For teleparallel case, we
examine that both hydrostatic and anisotropic forces show same
behavior but in opposite direction and thus balance each other. This
implies that the wormhole solution in teleparallel case satisfy
equilibrium condition. In case of power-law $f(T)$ function, this
condition shows wormhole solutions in equilibrium state as $r$
increases. For smaller values of $r$, these solutions do not satisfy
equilibrium condition properly or we may remark that these solutions
are less stable. For specific $b(r)$ function, forces
$\mathcal{F}_{hf}$ and $\mathcal{F}_{af}$ do not balance each other.
Since all the curves are in opposite manner having no similarity.
Therefore the physically acceptable wormhole solutions are not in
equilibrium form.

\section{Conclusion}

A wormhole represents shortcut distance to connect different regions
of the universe. To study these solutions, the violation of NEC
plays the key role which is associated with the exotic matter. To
minimize the usage of exotic matter is an important issue which
leads to explore a realistic model in favor of wormhole. In this
paper, we have studied static spherically symmetric wormhole
solutions in $f(T)$ gravity by taking NCL background. We have
developed the $f(T)$ field equations in terms of effective
energy-momentum tensor by taking non-diagonal tetrad and proved that
this tensor is responsible for the WEC violation. By imposing the
condition on matter content to thread the wormhole solutions, we
have assumed either the $f(T)$ or shape function and constructed the
other one. The graphical behavior of these solutions is discussed.

For power-law $f(T)$ model, we have discussed two cases:
teleparallel gravity ad $f(T)$ gravity in quadratic form. Both
models satisfied the conditions for wormhole geometry on shape
function. The WEC condition is violated for first case while
possibility of tiny wormhole solution is examined for $f(T)$ case.
In another case of particular form of power-law shape function, we
have analyzed the wormhole geometry. We have found physically
acceptable wormhole solutions as WEC satisfied in this case. Also,
we have examined the stability of these wormhole solutions with the
help of generalized Tolman-Oppenheimer-Volkov equation. We have
found that teleparallel NCL wormhole solutions are stable while
$f(T)$ NCL wormhole solutions are less stable. For the wormhole
solutions in specific shape function case, we have obtained unstable
solutions.

Bhar and Rahaman \cite{GR} investigated whether the wormhole
solutions exists in different dimensional non-commutative inspired
spacetimes with Lorentzian distribution. For four and five
dimensional spacetime, there exist wormhole solutions while no
solution for higher dimensions. They observed a stable wormhole,
i.e., satisfying flare out conditions and violating energy
conditions for four dimensions. It is interesting to note that for
underlying case, we also obtain asymptotically flat and stable
solutions in telaparallel case, i.e., the behavior of all conditions
is identical. The wormhole solutions \cite{11} in the background of
NC geometry give physically acceptable wormhole solutions in $f(T)$
gravity case while in teleparallel case, energy conditions violate
as per requirement. However this work is done taking diagonal tetrad
which is less in interest for spherical symmetry. In case of
Lorentzian distributed NC background, we have found physically
acceptable wormhole solutions in more stable form. We have used
non-diagonal tetrad which is the favorable choice for spherically
symmetry. The NC wormholes in $f(R)$ gravity with Lorentzian
distributed are discussed by Rahaman et al. \cite{frook}. They
studied the same cases and found violation of energy conditions in
all cases. That is no physically acceptable solutions in $f(R)$
gravity. However, we have found physically acceptable as well as
stable solutions.

\end{document}